\documentstyle[12pt,a4,cite,rotate,epsf]{article}
\newcommand{\cC}{{\cal C}}
\newcommand{\cD}{{\cal D}}

\newcommand{\ep}{\epsilon}
\newcommand{\gl}{\gamma_E}

\newcommand{\nn}{\nonumber}

\newcommand{\wt}{\widetilde}
\newcommand{\as}{\alpha_{s}}

\newcommand{\eqn}[1]{(\ref{#1})}

\newcommand{\gev}{\mbox{\rm GeV}}

\newcommand{\MSb}{{\overline{MS}}}

\newcommand{\smvs}{\vbox{\vskip 8mm}}
\newcommand{\bmvs}{\vbox{\vskip 10mm}}

\newcommand{\newsection}[1]{\section{#1}\setcounter{equation}{0}}

\begin{document}


\date{\small (September 1997)}

\author{
{\normalsize\bf Markus Eidem\"uller and Matthias Jamin} \\
\ \\
{\small\sl Institut f\"ur Theoretische Physik, Universit\"at Heidelberg,} \\
{\small\sl Philosophenweg 16, D-69120 Heidelberg, Germany}\\
}

\title{
{\small\sf
\rightline{HD-THEP-97-49}
\rightline{hep-ph/9709419}
}
\bigskip
\bigskip
{\Huge\bf QCD field strength correlator \\
at the next-to-leading order \\}
}

\maketitle
\thispagestyle{empty}

\begin{abstract}
\noindent
The gauge invariant two-point correlation function of the gauge field
strength tensor is calculated in perturbation theory at the next-to-leading
order. Besides a direct calculation in perturbative QCD we also present
a derivation of the correlation function in the heavy quark effective theory.
Our results are briefly compared with recent determinations of the field
strength correlator on the lattice.
\end{abstract}

\vspace{1cm}


\newpage
\setcounter{page}{1}


\newsection{Introduction}

The gauge invariant two-point correlation function of the QCD
field strength tensor $F^a_{\mu\nu}(x)$ in the adjoint representation
can be defined as
\begin{equation}
\label{eq:1.1}
\cD_{\mu\nu\lambda\omega}(z) \; \equiv \; \langle 0|T\{F^a_{\mu\nu}(y)
{\cal P}e^{\,gf^{abc}z^\tau\int_0^1 d\sigma A^c_\tau(x+\sigma z)}
F^b_{\lambda\omega}(x)\}|0\rangle
\end{equation}
where the field strength $F^a_{\mu\nu}=\partial_\mu A^a_\nu-\partial_\nu
A^a_\mu+gf^{abc}A^b_\mu A^c_\nu$, $z=y-x$ and ${\cal P}$ denotes path
ordering of the exponential. In general, the gauge invariant field strength
correlator could be defined with an arbitrary gauge string connecting the
end points $x$ and $y$. To simplify the calculation, however, in this
work we shall restrict ourselves to a straight line.

The field strength correlator $\cD_{\mu\nu\lambda\omega}(z)$ plays an
important role in non-perturbative approaches to QCD \cite{svz:79,vol:79,
leu:81,dos:94}. It is the basic quantity in the stochastic model of
the QCD vacuum \cite{dos:87,ds:88,sim:88} and in the description of high
energy hadron-hadron scattering \cite{nr:84,ln:87,kd:90,dfk:94}. In the
spectrum of heavy quark bound states it governs the effect of the gluon
condensate on the level splittings \cite{gro:82,cgo:86,sty:95} and it is
useful for the determination of the spin dependent parts in the heavy quark
potential \cite{sim:89}.

Until today, only the leading non-perturbative contribution to the
field strength correlator which is related to the gluon condensate
has been included in the phenomenological analyses. More details on this
subject as well as further references can be found in the review article
by Dosch \cite{dos:94}. However, as yet, higher order perturbative
corrections to the correlator $\cD_{\mu\nu\lambda\omega}(z)$ have only
been considered in the context of QED \cite{vz:89}. Actually in this
case it is possible to calculate the correlator to all orders in the
fine structure constant as a function of the QED $\beta$-function.
It is of clear interest for applications in non-abelian gauge theories
to also have control over the radiative corrections in the regime where
perturbation theory is applicable. 

Our paper is organised as follows. In section 2 results for the field
strength correlator up to next-to-leading order in perturbative QCD
will be presented and some details of the calculation are discussed.
An alternative derivation of the correlator from a hybrid quark gluon
current in the framework of the heavy quark effective theory (HQET)
is given in section 3. In section 4, we make a brief comparison of
our results with recent determinations of the field strength correlator
on the lattice \cite{gmp:97,egm:97} and summarise our work.

\begin{figure}[thb]
\centerline{
\epsfxsize=1.6in
\epsffile{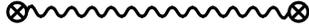} }
\caption[]{Leading order diagram for the field strength correlator.
\label{fig:1}}
\end{figure}

\newsection{Field strength correlator}

 From the Lorentz structure of the field strength correlator it follows
that the correlator can be parametrised in terms of two scalar functions
$\cD(z^2)$ and $\cD_1(z^2)$ \cite{dos:87,ds:88,sim:88}:
\begin{eqnarray}
\label{eq:2.1}
\cD_{\mu\nu\lambda\omega}(z) & = & \Big[\,g_{\mu\lambda}g_{\nu\omega}-
g_{\mu\omega}g_{\nu\lambda}\,\Big]\Big(\,\cD(z^2)+\cD_1(z^2)\,\Big) \nn \\
\smvs
& & \hspace{-3.9mm} +\,\Big[\,g_{\mu\lambda}z_\nu z_\omega-g_{\mu\omega}
z_\nu z_\lambda-g_{\nu\lambda} z_\mu z_\omega+g_{\nu\omega}z_\mu z_\lambda
\,\Big]\,\frac{\partial\cD_1(z^2)}{\partial z^2} \,.
\end{eqnarray}

At the leading order the functions $\cD^{(0)}(z^2)$ and $\cD_1^{(0)}(z^2)$
are readily calculated from the diagram of fig.~1. Note that at this order
the path ordered exponential in eq.~\eqn{eq:1.1} reduces to $\delta^{ab}$.
For an arbitrary gauge group $SU(N)$, we then find
\begin{eqnarray}
\label{eq:2.2}
\cD^{(0)}(z^2) & = & 0 \,, \nn \\
\smvs
\cD_1^{(0)}(z^2) & = & (N^2-1)\,\frac{\Gamma(2-\ep)}{\pi^{2-\ep}z^{4-2\ep}}
\; \stackrel{\ep\rightarrow 0}{=} \; (N^2-1)\,\frac{1}{\pi^2 z^4} \,.
\end{eqnarray}
Throughout this work, we use dimensional regularisation in $D=4-2\ep$
dimensions. In eq.~\eqn{eq:2.2} we have also given $\cD_1^{(0)}$ for
arbitrary $\ep$, because this result will be used for the renormalization
of the next-to-leading order. 

At the next-to-leading order, we have to calculate the diagrams of
figs.~2 and 3. The graphs of fig.~2 correspond to diagrams where the
string term does not contribute whereas in fig.~3 contractions with the
string, denoted by the dashed line, arise. The separate results
for all diagrams are presented in appendix~A. A complication arises
in the calculation of the diagram of fig. 3c). As can be seen from the
logarithm in eq.~\eqn{eq:a.5} this diagram generates a divergence, and
thus a logarithmic contribution to $D^{(1)}(z^2)$, although $D^{(0)}(z^2)$
vanishes and no immediate counterterm proportional to $D^{(0)}(z^2)$
is present. However, the vertex correction of a field strength tensor
times the string exponential generates new operators which, when
inserted in the two-point function, produces a mixing into $D^{(1)}(z^2)$.
For a proper renormalization of the correlation function this operator
mixing has to be taken into account.

\begin{figure}[htb]
\centerline{
\epsfxsize=4in
\epsffile{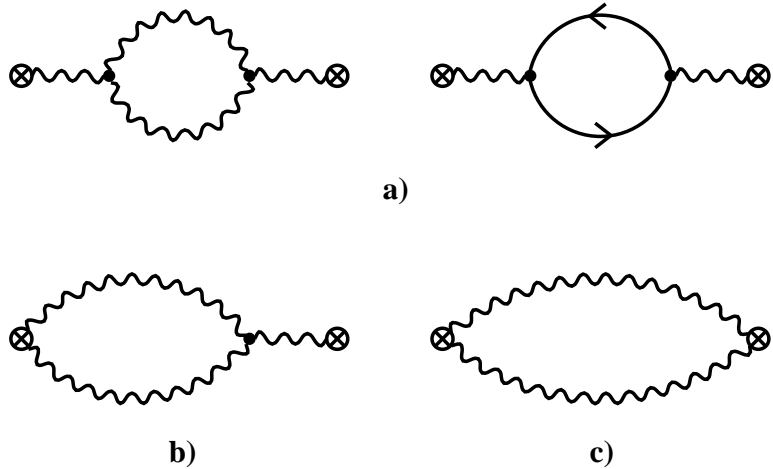} }
\caption[]{Next-to-leading order diagrams for the field strength correlator
without string contribution. The diagrams a) implicitly include the ghost
contribution. \label{fig:2}}
\end{figure}
\begin{figure}[hbt]
\centerline{
\epsfxsize=4in
\epsffile{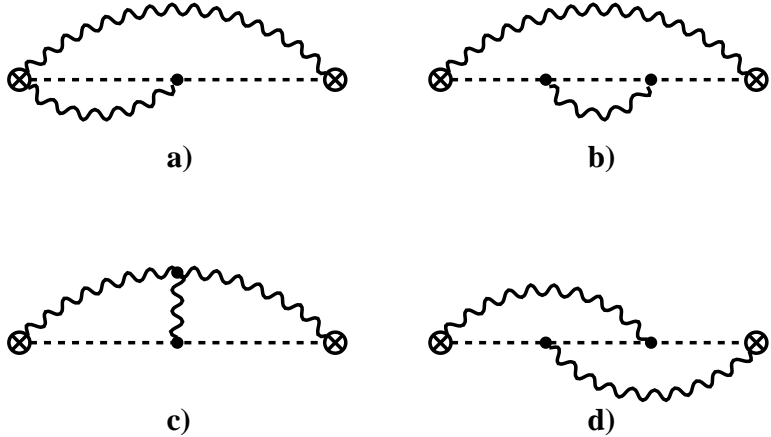} }
\caption[]{Next-to-leading order diagrams for the field strength correlator
with string contribution. \label{fig:3}}
\end{figure}

The final results for the perturbative next-to-leading order corrections
to the field strength correlator are then found to be
\begin{eqnarray}
D^{(1)}(z^2) & = & N(N^2-1)\,\frac{1}{\pi^2 z^4}\,\frac{\as}{\pi}\,
\biggl[\,-\,\frac{1}{4}\,L+\frac{3}{8}\,\biggr] \,, \label{eq:2.3} \\
\smvs
D_1^{(1)}(z^2) & = & N(N^2-1)\,\frac{1}{\pi^2 z^4}\,\frac{\as}{\pi}\,
\biggl[\,\biggl(\frac{\beta_1}{2N}-\frac{1}{4}\biggr)L+\frac{\beta_1}{3N}+
\frac{29}{24}+\frac{\pi^2}{3}\,\biggr] \,, \label{eq:2.4}
\end{eqnarray}
with $L=\ln(e^{2\gl}\mu^2 z^2/4)$, $\mu$ being a renormalization scale
in the $\MSb$ scheme and the first coefficient of the $\beta$-function
$\beta_1=(11N-2f)/6$.

 From the contributing diagrams it can be seen that $D_1^{(1)}(z^2)$
has a logarithmic contribution which is proportional to $\beta_1$ and
is related to the renormalization of the QCD coupling constant. This
logarithm can be resummed by considering
$\alpha_s(\mu^2) \cD_{\mu\nu\lambda\omega}(z)$ and choosing as the
renormalization scale in $\alpha_s$ $\mu^2=4e^{-2\gl}/z^2$. The
coefficients of the remaining logarithms are equal for $D^{(1)}(z^2)$
and $D_1^{(1)}(z^2)$ such that
\begin{equation}
\label{eq:2.5}
\alpha_s\Big(D_1(z^2)-D(z^2)\Big) \; = \; (N^2-1)\,\frac{\alpha_s(
4e^{-2\gl}/z^2)}{\pi^2 z^4}\,\biggl\{\,1+N\frac{\alpha_s}{\pi}\,\biggl[
\,\frac{\beta_1}{3N}+\frac{5}{6}+\frac{\pi^2}{3}\,\biggr]\,\biggr\}
\end{equation}
is a scale independent quantity at the next-to-leading order.

Even though the scale of $\alpha_s(\mu^2)$ in the next-to-leading order
term is only unambiguously fixed at the next-next-to-leading order, for
our numerical analysis let us assume $\mu^2\approx 1/z^2$ to be a natural
scale. From eq.~\eqn{eq:2.5} we obtain that at $0.2\,{\rm fm}$, which
corresponds to $1\,\gev$, the next-to-leading order correction is roughly
100\% and even at $4\,\gev$ or $0.05\,{\rm fm}$, the radiative correction
still amounts to about 50\% of the leading term. This finding entails that
the ${\cal O}(\alpha_s)$ correction to the field strength correlator is
large and that for phenomenological applications further understanding of
this large correction should be achieved.

\newsection{Field strength correlator from HQET}

The field strength correlation function $\cD_{\mu\nu\lambda\omega}(z)$
can also be calculated indirectly within the framework of the
HQET.\footnote{For a review on HQET as well as original references the
reader is referred to \cite{neu:94}.}
Consider the two-point correlator of the hybrid current
$(h^a F_{\mu\nu}^a)(x)$,
\begin{equation}
\label{eq:3.1}
\wt\cD_{\mu\nu\lambda\omega}(z) \; \equiv \; \langle 0|
T\{(h^a F^a_{\mu\nu})(y)\,(\bar h^b F^b_{\lambda\omega})(x)\}|0\rangle \, ,
\end{equation}
where $h^a(x)$ is an octet of heavy quark fields, interacting according
to QCD in the adjoint representation. From a path integral representation
of the field strength correlator it can be shown that \cite{eid:97}
\begin{equation}
\label{eq:3.2}
\wt\cD_{\mu\nu\lambda\omega}(z) \; = \; S(z)\,\cD_{\mu\nu\lambda\omega}(z) \,
\end{equation}
with $S(z)$ being the coordinate space propagator of a heavy quark field
defined by $T\{h^a(y)\bar h^b(x)\}=\delta^{ab} S(z)$.
In this derivation the heavy quark propagator replaces the path ordered
exponential and serves to make the correlation function gauge invariant.
A great advantage of this representation of the field strength correlator
is the fact that for its perturbative evaluation we only have to deal
with local operators.

The diagrams which have to be calculated in order to obtain
$\wt\cD_{\mu\nu\lambda\omega}(z)$ at the next-to-leading order are two-loop
diagrams similar to the diagrams of figs.~2 and 3. In the diagrams of fig.~2
there is an additional heavy quark propagator connecting the external
vertices and in the diagrams of fig.~3, the heavy quark propagator replaces
the gauge string. All integrals which were needed in the course of this
calculation were already known from HQET \cite{kil:94,abn:97}.\footnote{The
authors would like to thank M. Neubert for providing an integral needed
in the calculation of diagram 3c).} Analogously to the previous section,
in the case of diagram 3c), we have to take into account that under
renormalization the operator $(h^a F_{\mu\nu}^a)(x)$ mixes into new operators.

Calculating all diagrams and performing the renormalization diagram
by diagram, we found complete agreement with the direct QCD calculation
of the previous section and the results presented in appendix~A \cite{eid:97}.
This strong check gives us confidence in the correctness of our results.
In the next section we shall compare these results to most recent
determinations of the field strength correlation function on the lattice.

\newsection{Discussion}

Quite recently, the gluon field strength correlator
$\cD_{\mu\nu\lambda\omega}(z)$ has been measured on the lattice
\cite{gmp:97,egm:97}. The quantity determined in refs.~\cite{gmp:97,egm:97}
was
\begin{equation}
\label{eq:4.1}
\cC_{\mu\nu\lambda\omega}(z) \; = \; g^2 T_R\,
\cD_{\mu\nu\lambda\omega}(z)
\end{equation}
where in our conventions $T_R=1/2$
and the corresponding functions $\cC(z^2)$ and $\cC_1(z^2)$ are defined
analogously to eq.~\eqn{eq:2.1}. Our central results of eqs.~\eqn{eq:2.3}
and \eqn{eq:2.4} for the field strength correlator show that
it generally depends on the renormalization scale as well as the
renormalization scheme. A direct comparison to the lattice results of
refs.~\cite{gmp:97,egm:97} is therefore difficult and a quantitative
analysis would require a similar calculation of $\cC_{\mu\nu\lambda\omega}(z)$
in lattice perturbation theory to establish the relation between the two
schemes. It is to be expected that the scheme dependence goes beyond
merely replacing $\alpha_s^{\MSb}$ by the lattice coupling constant and
that there should be additional constant terms. Nevertheless, we shall
attempt to present some qualitative observations.

In ref.~\cite{gmp:97} the pure QCD correlator has been measured for four
values of $\beta=6/g^2$ between $6.6$ and $7.2$. It was found that an
acceptable fit to $\cC_1(z^2)$ can be obtained by a pure perturbative
$1/z^4$ behaviour. Dividing out $g^2 T_R$ with an average value $g^2=0.87$,
we then obtain $\cD_1^{lat}(z^2)\approx 0.69/z^4$. This can be compared
with the leading order result of eq.~\eqn{eq:2.2},
$\cD_1^{(0)}(z^2)\approx 0.81/z^4$ showing reasonable agreement at the 20\%
level. However, whereas for a natural $\MSb$ scale $\mu^2\approx 1/z^2$
the next-to-leading order correction to $\cD_1(z^2)$ turns out to be
large and positive, the lattice results suggest a moderate, negative
correction from higher orders. On the other hand in perturbation theory
$\cD(z^2)$ vanishes at the leading order and the next-to-leading order
correction in the $\MSb$ scheme is found to be much smaller than the
correction to $\cD_1(z^2)$. Surprisingly, on the lattice $\cC(z^2)$ was found
to be larger than $\cC_1(z^2)$. Hopefully, a perturbative calculation in
the lattice regularisation scheme will clarify these issues in the future.

In QED the gauge string exponential is absent and higher order corrections
are only due to the renormalization of the electric charge. Vainshtein and
Zakharov calculated the quantity $\cD_{\mu\nu}^{\;\;\mu\nu}(z^2)$ in terms
of the QED $\beta$-function \cite{vz:89}. However, the leading term for
$\cD_{\mu\nu}^{\;\;\mu\nu}(z^2)$ in QED starts at order $\alpha^2$. Hence
it is not possible to compare our results at the next-to-leading order
to the QED case presented in ref.~\cite{vz:89}.

To summarise, in this work we have presented a calculation of the gauge
invariant field strength correlator at the next-to-leading order in
perturbation theory. It was found that the correlator depends on both
the renormalization scale as well as the renormalization scheme.
Constructing a scale independent quantity at this order, it turned out
that the first order correction is large, namely roughly 50\% to 100\%
for distances of $z=0.05\,{\rm fm}$ to $0.2\,{\rm fm}$. Because of the
scheme dependence of the field strength correlator, for a sound comparison
with recent lattice data, an analogous calculation of the correlator in
lattice perturbation theory would be required. We hope to return to this
questions and phenomenological applications of this work in the future.

\vspace{6mm} \noindent
{\Large\bf Acknowledgments}
 
\vspace{3mm} \noindent
The authors gratefully acknowledge collaboration with H.~G.~Dosch in
early stages of this work. We also thank A.~Di~Giacomo, W.~Kilian,
M.~Neubert, V.~Shevchenko and Yu.~A.~Simonov for helpful discussions.

\newpage
\appendix{\LARGE\bf Appendix}

\newsection{Results for $D^{(1)}$ and $D_1^{(1)}$}

Below, we shall present our results for all diagrams separately. In order
to check explicitely the gauge invariance of the field strength correlator,
we have performed the calculation in a general covariant gauge.
The next-to-leading order corrections to the functions $D(z^2)$ and
$D_1(z^2)$ can be written as follows:
\begin{equation}
\label{eq:a.1}
  D^{(1)}(z^2) \; = \; D_1^{(0)}(z^2)\,  G^{(1)}(z^2) \quad \hbox{and} \quad
D_1^{(1)}(z^2) \; = \; D_1^{(0)}(z^2)\,G_1^{(1)}(z^2) \,,
\end{equation}
with $D_1^{(0)}(z^2)$ given in eq.~\eqn{eq:2.2}.

For the separate renormalised contributions to $G^{(1)}(z^2)$ we find:
\begin{eqnarray}
G_{2c}^{(1)} & = & N\,\frac{\as}{\pi}\,\biggl[\,\frac{1}{8}+
\frac{a}{8}\,\biggr] \,, \label{eq:a.2} \\
\bmvs
G_{3a}^{(1)} & = & N\,\frac{\as}{\pi}\,\biggl[\,\frac{3}{4}-
\frac{a}{4}\,\biggr] \,, \label{eq:a.3} \\
\smvs
G_{3b}^{(1)} & = & N\,\frac{\as}{\pi}\,\biggl[\,-\,\frac{3}{8}+
\frac{a}{8}\,\biggr] \,, \label{eq:a.4} \\
\smvs
G_{3c}^{(1)} & = & N\,\frac{\as}{\pi}\,\biggl[\,-\,\frac{1}{4}\,L
-\frac{1}{8}\,\biggr] \,, \label{eq:a.5}
\end{eqnarray}
where $L=\ln(\pi e^{\gl}\nu^2 z^2)$ and $\nu^2$ is a renormalization scale in
the $MS$-scheme \cite{thv:72}. We have already subtracted the corresponding
counterterms diagram by diagram. To relate our expressions to the more
conventional $\MSb$-scheme \cite{bbdm:78} we have to use
$\nu^2=e^{\gl}\mu^2/(4\pi)$ where now $\mu^2$ is a renormalization scale in
the $\MSb$-scheme. All diagrams which have not been listed explicitely give
a vanishing contribution. Summing all diagrams, we obtain:
\begin{equation}
\label{eq:a.6}
G^{(1)}(z^2) \; = \; N\,\frac{\as}{\pi}\,\biggl[\,-\,\frac{1}{4}\,L
+\frac{3}{8}\,\biggr] \,.
\end{equation}

The separate contributions to $G_1^{(1)}(z^2)$ are found to be:
\begin{eqnarray}
G_{1,2a}^{(1)} & = & N\,\frac{\as}{\pi}\,\biggl[\,\biggl(\frac{\beta_1}{2N}-
\frac{3}{8}-\frac{a}{8}\biggr)L+\frac{\beta_1}{3N}-\frac{23}{48}+\frac{a}{4}
+\frac{a^2}{16}\,\biggr] \,, \label{eq:a.7} \\
\smvs
G_{1,2b}^{(1)} & = & N\,\frac{\as}{\pi}\,\biggl[\,\biggl(-\,\frac{5}{8}-
\frac{a}{8}\biggr)L-\frac{1}{4}-\frac{3a}{8}-\frac{a^2}{8}\,\biggr]
\,, \label{eq:a.8} \\
\smvs
G_{1,2c}^{(1)} & = & N\,\frac{\as}{\pi}\,\biggl[\,-\,\frac{1}{16}+
\frac{a^2}{16}\,\biggr] \,, \label{eq:a.9} \\
\bmvs
G_{1,3a}^{(1)} & = & N\,\frac{\as}{\pi}\,\biggl[\,-\,\frac{3}{4}+
\frac{a}{4}\,\biggr] \,, \label{eq:a.10} \\
\smvs
G_{1,3b}^{(1)} & = & N\,\frac{\as}{\pi}\,\biggl[\,\biggl(\frac{3}{4}-
\frac{a}{4}\biggr)L+\frac{11}{8}-\frac{a}{8}\,\biggr]
\,, \label{eq:a.11} \\
\smvs
G_{1,3c}^{(1)} & = & N\,\frac{\as}{\pi}\,\biggl[\,\frac{a}{2}\,L+
\frac{11}{8}+\frac{\pi^2}{3}\,\biggr]
\,. \label{eq:a.12}
\end{eqnarray}
The flavour dependent contributions have been rewritten in terms of
the first coefficient of the QCD $\beta$-function, $\beta_1=(11N-2f)/6$.
Again summing all diagrams, we obtain:
\begin{equation}
\label{eq:a.13}
G_1^{(1)}(z^2) \; = \; N\,\frac{\as}{\pi}\,\biggl[\,\biggl(\frac{\beta_1}{2N}-
\frac{1}{4}\biggr)L+\frac{\beta_1}{3N}+\frac{29}{24}+\frac{\pi^2}{3}\,\biggr]
\,.
\end{equation}

\newpage

\end{document}